\documentclass[bibyear]{aa}
\usepackage{times}
\usepackage{amsmath,mathrsfs,amssymb,amsbsy}
\usepackage{graphicx}
\usepackage{subfigure}
\usepackage{paralist}
\usepackage{color}
\usepackage{natbib}
\usepackage{aas_macros}
\usepackage{float}
\usepackage[colorlinks=true,linkcolor=blue, citecolor=blue]{hyperref}%
\graphicspath{{./Plots/}}
\usepackage{grffile} 

\newcommand{\kms}{\,km\,s$^{-1}$} 

\usepackage{gensymb} 
\usepackage{soul} 

\newcommand{\teff}{$T_{\rm eff}$}
\newcommand{\logg}{$\log g$}
\newcommand{\meta}{\rm{[M/H]}}
\newcommand{\metaism}{$\rm{[M/H]_{ISM}}$}

\newcommand{\feh}{\rm {[Fe/H]}}

\def\Rsun{$R_{\odot}$}

\def\zmax{$Z_{\rm max}$}
\def\kms{\,{\rm km~s^{-1}}}
\def\kpc{\,{\rm kpc}}

\def\dex{\,{\rm dex}}
\def\Gyr{\,{\rm Gyr}}
\def\vphi{V_\phi}

\def\DeltaISM{\Delta\meta_{\rm ISM}}
\def\ltsima{$\; \buildrel < \over \sim \;$}
\def\simlt{\lower.5ex\hbox{\ltsima}}
\def\gtsima{$\; \buildrel > \over \sim \;$}
\def\simgt{\lower.5ex\hbox{\gtsima}}

\def\ltsima{$\; \buildrel < \over \sim \;$}
\def\simlt{\lower.5ex\hbox{\ltsima}}
\def\gtsima{$\; \buildrel > \over \sim \;$}
\def\simgt{\lower.5ex\hbox{\gtsima}}

\providecommand{\orcit}[1]{\protect\href{https://orcid.org/#1}{\protect\includegraphics[width=8pt]{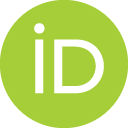}}}
    
\begin{document}

\title{First large scale spatial and velocity patterns \\ of local metal-rich stars in the Milky Way}
\author{
	G.~Kordopatis\orcit{0000-0002-9035-3920},\inst{\ref{oca}}\thanks{\email{georges.kordopatis@oca.eu}}
	D.~Feuillet\orcit{0000-0002-3101-5921},\inst{\ref{lund}, \ref{uppsala}} 
	C. Lehmann\orcit{0000-0002-1050-4400},\inst{\ref{lund}}
	S. Feltzing\orcit{0000-0002-7539-1638},\inst{\ref{lund}} 
	I. Minchev\orcit{0000-0002-5627-0355},\inst{\ref{aip}}
	V.~Hill\orcit{0000-0002-3795-0005},\inst{\ref{oca}}
	 H.~Ernandes\orcit{0000-0001-6541-1933}\inst{\ref{lund}} 
	}
\institute{
	Universit\'e C\^ote d'Azur, Observatoire de la C\^ote d'Azur, CNRS, Laboratoire Lagrange, Nice, France\label{oca}
	\and
	Lund Observatory, Department of Geology, Sölvegatan 12, SE-22362 Lund, Sweden\label{lund}
	\and
	Observational Astrophysics, Department of Physics and Astronomy, Uppsala University, Box 516, SE-751 20 Uppsala, Sweden\label{uppsala}
	\and
	Leibniz-Institüt für Astrophysik Potsdam (AIP), An der Sternwarte 16, D-14482, Potsdam, Germany\label{aip}
	}
\abstract
{
The present-day spatial and kinematic distribution of stars in the Milky Way provides key constraints on its internal dynamics and evolutionary history.
However,  identifying the correct tracers that highlight the mechanisms is far from being straightforward. The best probes are stars that stand out in terms of kinematics, chemistry or age, compared to the underlying population where they belong to. }
 { We aim at putting constraints on stellar radial migration and, in particular, to study observationally the effect of the latter on the disc dynamical heating.  }
{We select, throughout the Milky Way, stars that are more metal-rich than the interstellar medium  (ISM) at their guiding radius  (the so-called Local Metal-Rich stars, LMR) and investigate their chemo-kinematics. 
Until recently, existing catalogues did not contain such targets in large quantities, but one can now select  many millions of them  by using Gaia  photometric metallicities. Once selected, we investigate their kinematics and age distributions across the disc, and compare them to the stellar populations having the metallicity of the ISM.}
{ 
Compared to locally born stars with metallicities equal to the ISM's one, we find that LMR stars, at a given location, are always older (mean age up to 2\Gyr~older) and with velocity dispersions similar or slightly higher.
Furthermore, at a given metallicity, LMR stars are older at larger galactocentric radiii, reflecting the fact that LMR stars need time to migrate.
Finally, whereas we do not find any correlation between the location of the spiral arms and the spatial density of LMR stars, we find that the mean stellar eccentricity and mean ages show smaller values where the spiral arms are.
} 
{Our results   confirm a well established theoretical result that has not yet been formally confirmed via observations on large datasets without modelling:  churning is not  significantly heating the Galactic disc. Furthermore, the age distribution of these stars rule-out any significant contribution from Galactic fountains as their origin, and confirm the effect of the spiral arms on them. 
Although no clear signature of the Galactic bar is detected, its phase-mixed and diffuse influence --especially through interactions with spiral arms-- cannot be excluded. }


\titlerunning{Local metal rich stars  in Gaia}
\authorrunning{G.~Kordopatis et al.}

\maketitle

\section{Introduction}
\label{sec:Introduction}
Stellar population kinematics bring valuable insight into the dynamical processes that take place throughout the history of a galaxy.  In particular, stars in galactic discs are thought to be born with nearly circular orbits, broadly following the kinematics of the gas from which they were formed \citep[e.g.][]{Spitzer51, Wielen77}. Over time, however, several processes can affect their motion, such as accretion events \citep[e.g.][]{Velazquez99, Kazantzidis08, Gomez13, Laporte22}, interactions with giant molecular clouds \citep{Spitzer53, Lacey84, Ida93} and resonances with non axi-symmetries in the galactic potential,  such as the bar \citep[e.g.][]{DiMatteo13, Khoperskov20, Haywood24}, the spiral arms \citep[e.g][]{Minchev06, Schonrich09b} or overlapping patterns of both \citep[e.g][]{Quillen03, Minchev10, Daniel19, Marques25}.  

At the Lindblad resonances, stars increase their eccentricity while keeping their guiding radius unchanged \citep[e.g.][]{Lynden-Bell72}. 
Such kinematically hot stars create an effect known as  \emph{blurring}, which is causing stars to momentarily visit regions of the galaxy far from their  guiding
radius, with velocities lower than the one of the local standard of rest (LSR)  if they are at their apocentre or higher if they are at their pericentre. 
Resonances that occur at co-rotation, tend to change the angular momentum and hence the guiding radius, without (much) affecting the star's radial action \citep{Sellwood02}. This process is commonly called \emph{churning}. The location at which co-rotation resonances  occur depend on the pattern speed  (and hence on the strength)  of the considered structures and can change over time depending on the temporal evolution of the non axi-symmetries \citep{Chiba21, Khoperskov22, Sellwood22, CLi24} or the  transient mass clumps provoked by the overlapping long-lived structures with different pattern speeds \citep[][]{Marques25}.  

Whereas dynamical heating, which causes blurring, has been a well established mechanism in the Milky Way since the 1970s, identified to be mostly responsible for the increase of the age-velocity dispersion relation \citep[e.g.][]{Wielen77}, first direct observational evidence of  churning has been provided only a decade ago by  \citet{Kordopatis15a}. 
 In that paper, the authors selected   from the medium-resolution RAVE DR4 catalogue \citep{Kordopatis13b} stars with metallicities above solar in the extended solar neighbourhood (the so-called Super Metal- Rich stars, SMR, \citealt{Grenon72}), and found that they almost all had  eccentricities below $\sim0.2$, i.e. they could not just be visitors from the inner Galaxy on high eccentricity orbits.
Since these stars had metallicities above the metallicity of the local interstellar medium (ISM), they could not have been born locally either, unless they were formed from locally enriched material coming, for example, from Galactic fountains \citep[e.g.][]{Fraternali13, Grand19}.  The presence of such stars, has  been recently re-confirmed on a larger volume by, e.g., \citet{Lehmann24} using Gaia DR3 kinematics and APOGEE DR17 metallicities and ages.

The efficiency of churning and its effect on the disc has since then been tested by several studies, either by selecting explicitly SMR stars \citep[e.g.][]{Santos-Peral21, Dantas23, Lehmann24}, by determining the birth radii of the stars \citep[e.g.][]{Minchev18, Ratcliffe23b, Lu24}, by modelling the metallicity distribution function across the disc \citep[e.g.][]{Lian22}, or by forward-modelling the orbit evolution for stars of a given  age and metallicity and fitting it to data \citep{Frankel20}.  All of these studies were based on spectroscopic surveys that overall only contained few hundreds of thousands of stars, a number which drops significantly when probing large distances from the Sun or the tails of the metallicity distribution function. 

Nowadays, it has  been collegially accepted, although with rather indirect way and low statistics \citep[e.g.][]{Mackereth19, Frankel20, Sharma21}, that churning does not dynamically heat the Galactic  disc. That is, stars that migrate outwards via co-rotation resonances reach a given position with similar eccentricity and maximum distance above the Galactic plane (\zmax)
as locally born stars, due to conservation of the radial and vertical action \citep[][]{Minchev12c, Halle15, Sellwood25}. 
Indeed, as stars move outwards, the Galactic potential decreases, and in order to keep the action constant, one therefore needs to decrease the vertical velocity dispersion of the considered population  (see, however, \citealt{Hamilton24} and \citealt{Sellwood25} on the spiral theories and mechanisms that could cause this).

With the advent of Gaia's third data release \citep[Gaia DR3,][]{Gaia,GaiaDR3}, exquisite 3D kinematics of tens of millions of stars have become available \citep{Kordopatis23a} to investigate these claims more thoroughly. Together with these kinematics,  photometric metallicities  have become available for roughly a hundred million stars,  thanks to state-of-the-art machine learning techniques \citep{Andrae23, JLi24, Hattori24} and/or synthesis of cleverly chosen and calibrated photometric bands \citep[][]{Martin24}. Such photometric metallicities overcome problems of spectroscopy allowing for datasets with much simpler  selection effects and with homogeneously determined parameters up to relatively faint magnitudes and hence large distances. The synergy of the kinematic and metallicity catalogues therefore allows us to probe in a unique way rare stellar populations.

In this paper, we address the effect of churning on the disc structure using an unprecedentedly  large sample of more than $1.6\cdot10^6$ Local Metal-Rich stars (LMR) over a large volume,  with Gaia-DR3 kinematics and new age determinations. 
LMR stars are defined as having metallicities which are higher than the one of today's interstellar medium (ISM) at their guiding radius, and  are therefore candidates of  not being formed locally (unless formed from  in-falling metal-rich material, such as the Galactic fountains).
We define the ISM's metallicity adopting the \citet{daSilva23} metallicity $\log$-gradient, namely:  $\feh_{\rm ISM}=-0.907\cdot \log(R)+0.81$, as evaluated using classical Cepheids and open clusters. We note that adopting other gradients, such as the linear gradient of \citet{daSilva23}, or the one of \citet{Genovali14}, did not change significantly the results of this paper. 

\smallskip 

Section~\ref{sec:Data_selection} describes the dataset used and the quality filters applied. Section~\ref{sec:results} presents the results obtained for the overall kinematics and age distributions of LMR stars, and Sect.~\ref{sec:effect_of_spirals} describes how these results correlate with the presence of the spiral arms. Finally, Sect.~\ref{sec:conclusions} concludes and summarises our results.

\section{Data selection}
\label{sec:Data_selection}
The underlying sample that we are using is the union of the  Gaia's DR3 RVS sample \citep[$\sim 33\cdot10^6$ stars,][]{Katz23},  with the  golden sample of red giant branch (RGB) stars sample of \citet{Andrae23} ($\sim 17\cdot10^6$ stars, dubbed XGBoost). The latter has photometric metallicities derived combining the {\tt CatWISE} photometry together with the Gaia XP spectra and astrometry, trained adopting the labels from APOGEE DR17 \citep{Abdurrouf22} and for very metal-poor stars with the labels from \citet{Li22}. 

Below, we describe the different cuts we apply to this dataset. Figure~\ref{fig:quality_cuts}, in the Appendix, summarises their successive effect on the resulting metallicity distribution.

\subsection{Filtering based on Gaia astrometry and agreement with other photometric metallicities}
\label{subsec:ruwe_varpi_deltamh_selection}

\begin{figure}[tbp]
\centering
\includegraphics[width=\linewidth]{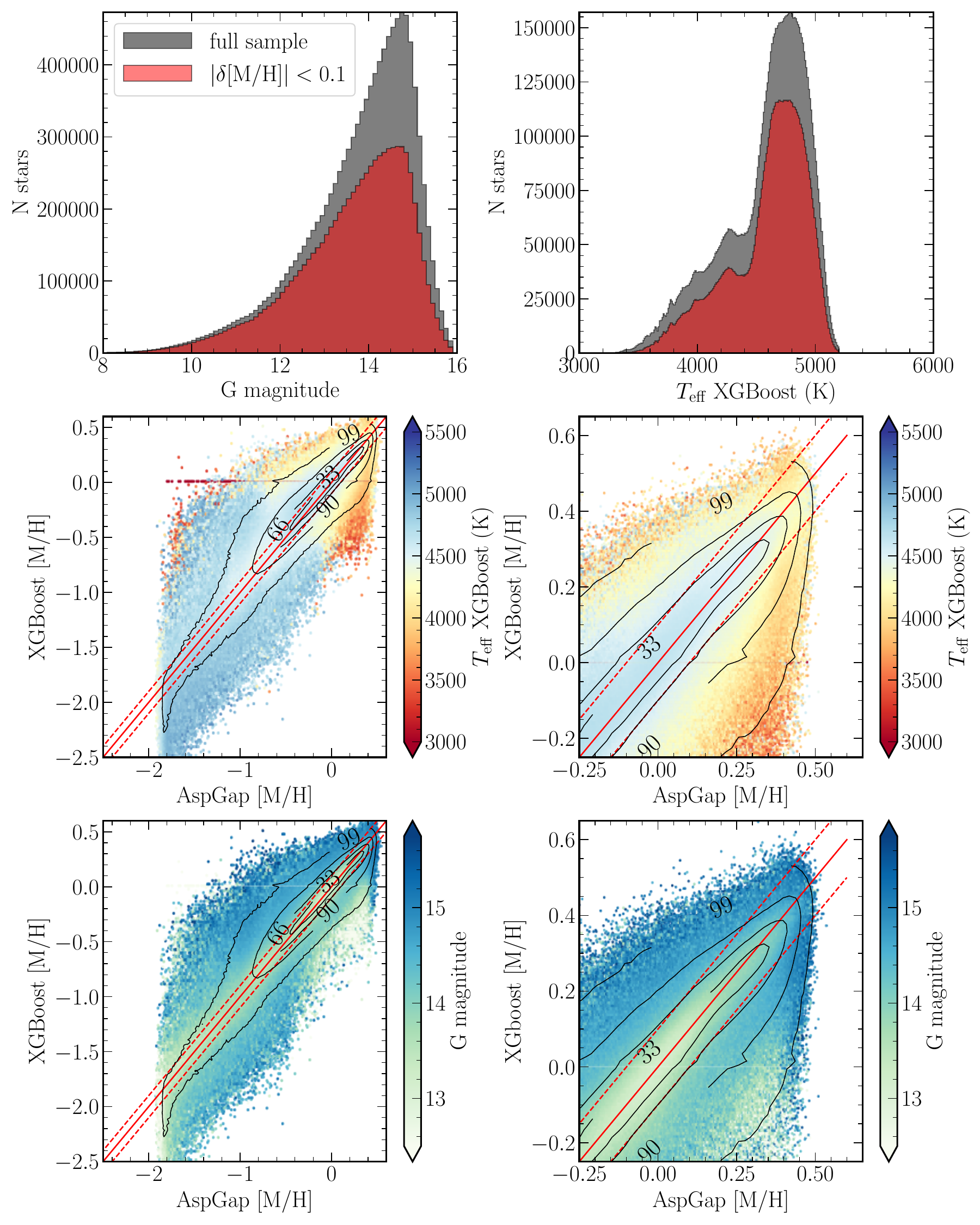}
\caption{\emph{Top:} G-magnitude distributions (left) and \teff~ distributions of the total XGBoost sample cross-matched with AspGap (in grey), and of the sub-sample for which differences in metallicity are smaller than 0.1\dex~(in red), see Sect.~\ref{sec:Data_selection}. 
\emph{Middle:} comparison between the XGBoost and the AspGap metallicities, colour-coded by \teff. The plot on the right zooms-in on the high-metallicity regime. \emph{Bottom: } Same as the middle plots, colour-coded by G magnitude. Black contour-lines represent 33, 66, 90 and 99 per cent of the sample. The red solid line represents the 1:1 relation, whereas the dashed ones are shifted by $\pm0.1$\dex~ from identity.   }
\label{fig:xgboost_aspgap_feh}
\end{figure}

In order to keep the stars with the most accurate metallicity estimates, we cross-match our sample with the \citet{JLi24}  catalogue (dubbed AspGap). The latter also provides  photometric metallicities together with global $\alpha$-abundances  for $37 \cdot 10^6$ RGB stars, derived solely based on the Gaia XP spectra and trained on both the APOGEE labels and spectra. 
  From the resulting cross-match, we arbitrarily select only the stars that have metallicities that agree within $0.1\dex$ between the two catalogues (see Fig.~\ref{fig:xgboost_aspgap_feh}). This corresponds roughly to 69 per cent of the cross-matched sample ($\sim$ 8.3 million stars out of the 11.9 million), mostly removing  stars fainter than $G=13.5$\,mag.

We further filter the stars based on the {\tt RUWE} parameter (imposing $<1.4$, \citealt{Lindegren_RUWE})  in order to minimise the possible contamination from variable and/or multiple stars, and on the parallax uncertainty (adopting $\sigma_\varpi / \varpi <$0.2\footnote{Stricter cuts were explored, without changing the results of the paper.}) to  avoid having positions (and hence kinematics) being strongly dependent on the distance-prior of the stars \citep[see, e.g.][]{Bailer-Jones21}.  After these extra cuts, the sample consists of 7.68 million stars. 

\begin{figure}[tbp]
\centering
\includegraphics[width=\linewidth]{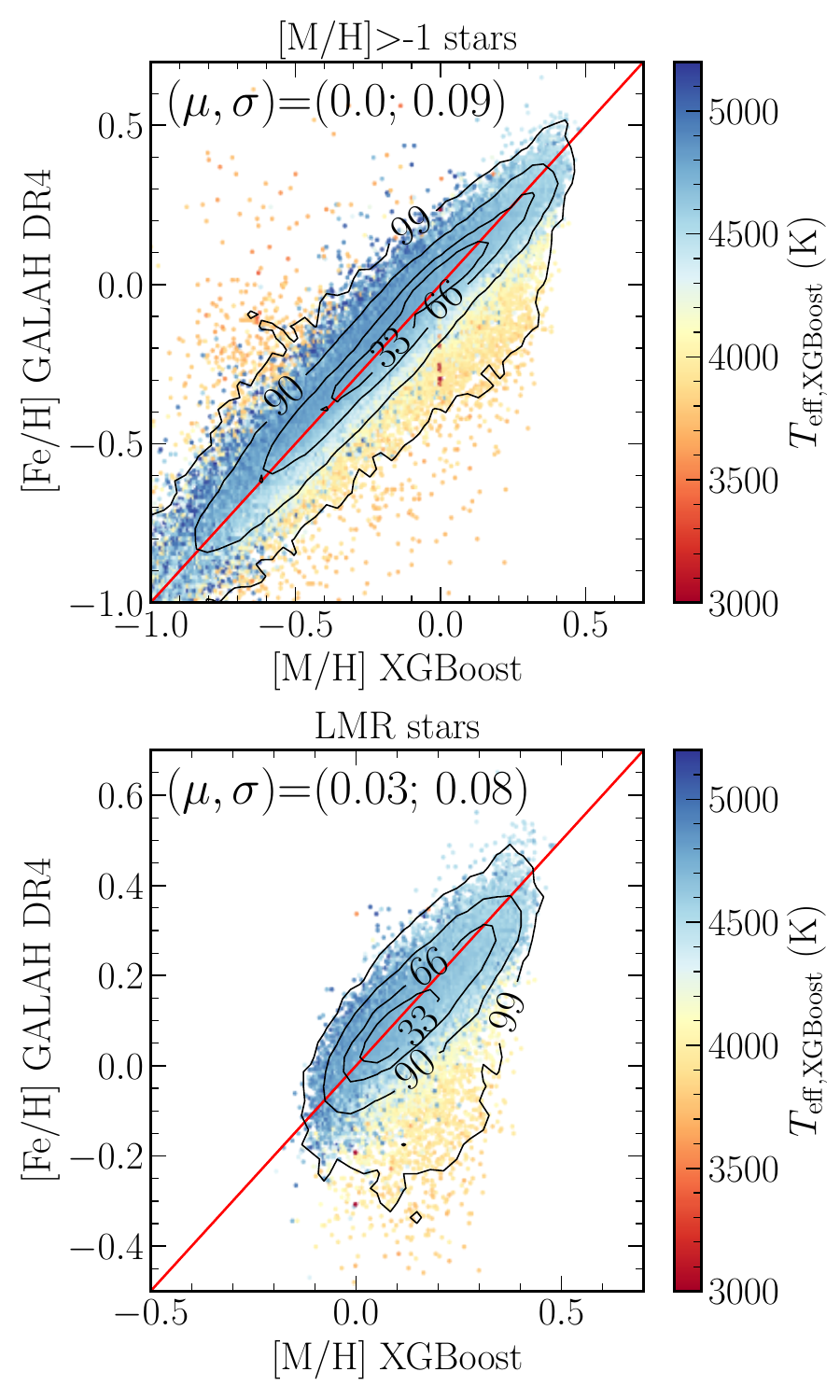} 
\caption{Comparison of XGboost metallicities with GALAH DR4 iron abundances.  All of the panels include the selection described in Sect.~\ref{subsec:ruwe_varpi_deltamh_selection} as well the {\tt flag\_sp==0} selection from GALAH. The colour-code is XGboost \teff, and the numbers within each plot indicate the mean offset and dispersion of the difference. The top plot considers all of the stars with $\meta>-1$, whereas the bottom plot  considers only the local metal-rich stars, i.e. stars with metallicities above the ISM's one at their guiding radius.  }
\label{fig:xgboost_galahdr4_feh}
\end{figure}

\subsection{Filtering on effective temperature}
\label{sec:teff_filter}

A careful investigation of the middle plots of Fig.~\ref{fig:xgboost_aspgap_feh} shows that very cool stars (in red) may need some further filtering to ensure a high reliability of the metallicities of the sample. 
To assess whether this is the case or not, we compare the selection described in the previous section with the purely spectroscopic metallicities\footnote{The suggested filtering {\tt flag\_sp==0} has been applied.} from GALAH DR4 \citep{Buder25}.
The comparison, shown in Fig.~\ref{fig:xgboost_galahdr4_feh} (top: all of the stars with $\meta>-1$, bottom: just the LMR stars), confirms that the most discrepant stars are found to be the ones with the lowest temperatures (\teff$_{,\rm XGBoost} \lesssim 4250$\,K). 

It is beyond the scope of this paper to assess which among XGBoost or GALAH DR4 is closer to reality \citep[see, however,][]{Kos25}. On the one hand, cool metal-rich stars are notoriously difficult to parameterise spectroscopically due to the multiple presence of significant absorption features formed by molecules making  the continuum placement difficult \citep[e.g.][]{Kordopatis11a, Kordopatis23b}. On the other hand, the wider wavelength coverage of the XP spectra may, in principle, lead to parameters with smaller biases than the ones obtained spectroscopically, despite the lower resolving power. However, XGBoost parameters have been obtained with a training  set whose parameters are also derived spectroscopically, and photometric parameters may also be prone to degeneracies with interstellar reddening. Given these points, we therefore decide to keep only the stars in the \teff~regime where there is a good agreement with GALAH, i.e. \teff$_{,\rm XGBoost}>4250$\,K. This filtering  removes $\sim 1.3 \cdot 10^6$ additional stars.
We note that once all of the filters are applied to the data (see Fig.~\ref{fig:quality_cuts}), XGBoost's \teff~ are in very good agreement with AspGap ones. As a matter of fact,  the  distributions of the difference between the two temperature estimates have a mean of less than 10\,K and a dispersion smaller than 60\,K, for the LMR subsample and the full (i.e. without a metallicity-cut) one.

For the remainder of the paper, unless specified otherwise, $\meta$ will be referring to the XGBoost metallicity.

\subsection{Adopted positions, kinematics and orbital parameters}
We select the positions (Galactocentric cylindrical $R$, and cartesian $X, Y, Z$), kinematics (cylindrical $V_R, V_\phi, V_Z$) and orbital parameters (eccentricity $e$, maximum distance from the Galactic plane \zmax) from \citet{Kordopatis23a}, adopting the set computed with the geometric line-of-sight distances of \citet{Bailer-Jones21}. The assumed values for the Sun's position and velocities are $(R,Z)_\odot=(8.249,0.0208)\kpc$, $(V_R, \vphi, V_Z)_\odot=(-9.5, 250.7, 8.56)\kms$ \citep{Gravity20, Bennett19, Reid20}. 
We also evaluate the guiding radius of the stars, $R_{\rm guide}= R \cdot \vphi/v_{\rm lsr}$ where $v_{\rm lsr}=238.5\kms$ \citep{Schonrich10}.

\begin{figure*}[tbp]
\begin{center}
\includegraphics[width=0.49\linewidth]{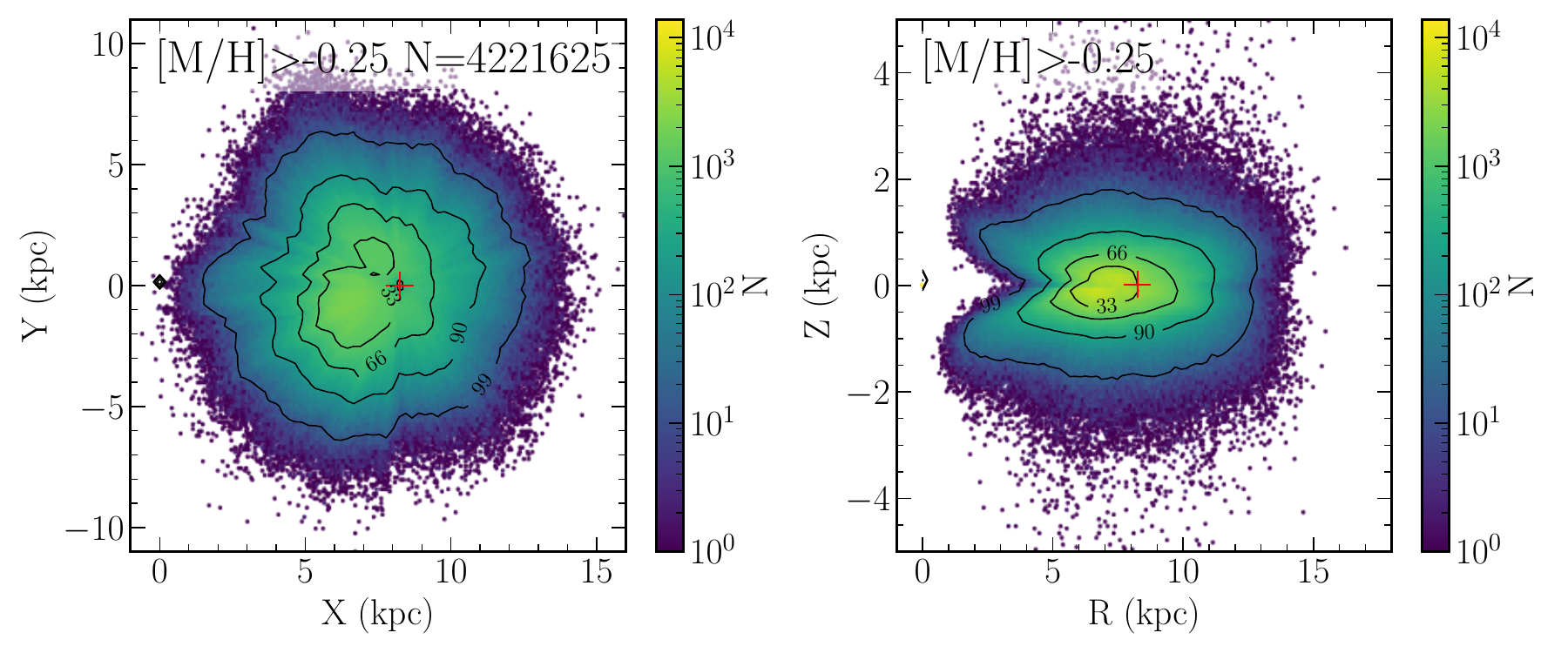} 
\includegraphics[width=0.49\linewidth]{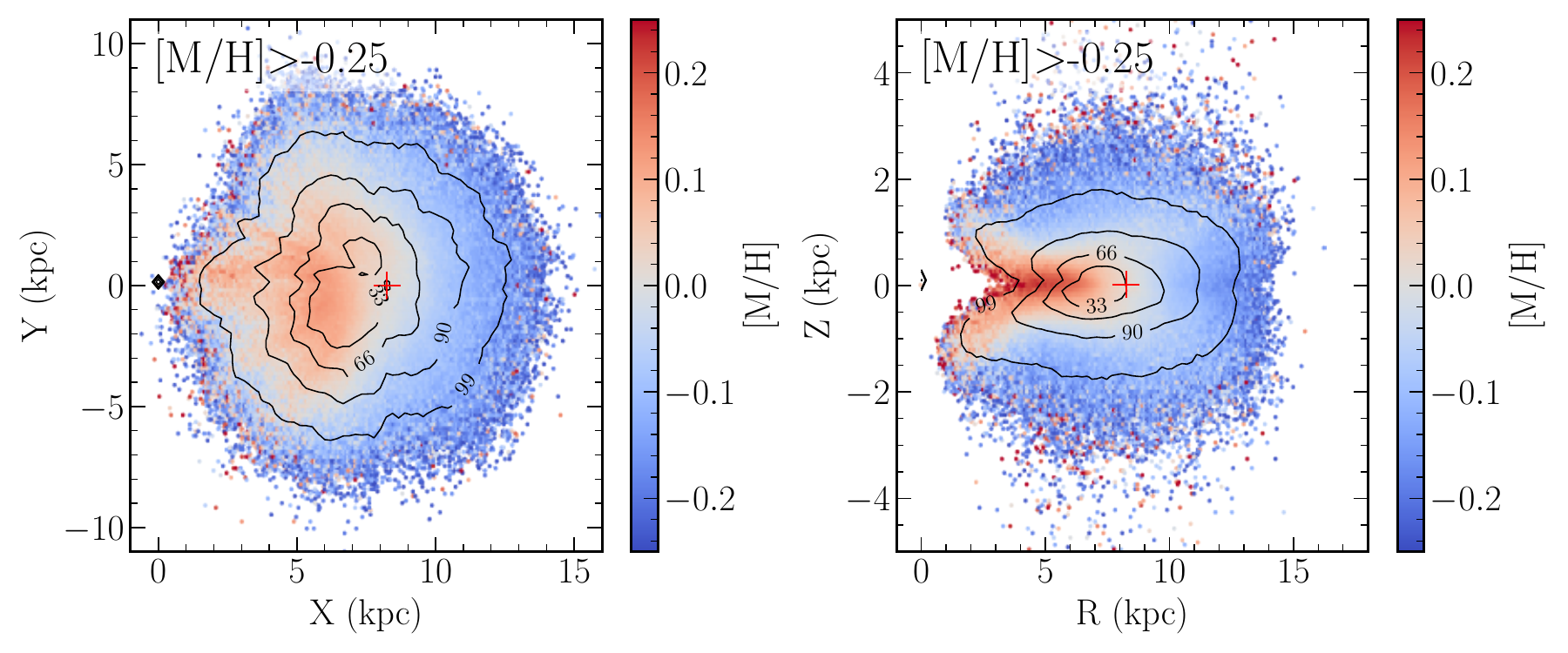}
\caption{Spatial distribution of stars with $\meta>-0.25$  in face-on view (X-Y, panels 1 and 3) and edge-on view (R-Z, panels 2 and 4) after the selections described in Sect.~\ref{sec:Data_selection}.  The position of the Sun is indicated by a red `+' sign, at $R=8.249\kpc$.   The two plots on the left are colour-coded by number of stars whereas the ones on the right by average metallicity. Iso-contour lines containing 33, 66, 90 and 99 per cent of the distribution are plotted inside each panel.}
\label{fig:XY_global}
\end{center}
\end{figure*}

 Figure~\ref{fig:XY_global} shows the spatial distribution (top-view and side-view) of all of the stars with $\meta>-0.25$, colour-coded by density (panels on the left), and metallicity (panels on the right). 
  Globally, our sample covers smoothly the $X-Y$ and $R-Z$ plane, avoiding the mid-plane for Galactocentric distances smaller than $R\sim4\kpc$, i.e. towards the Galactic centre. We also note that that there is a lower density of stars  at the lowest |Z| distances, due to the various selection functions of the RVS, XGBoost and AspGap samples.
In Fig.~\ref{fig:XY_global}, a clear radial metallicity gradient is seen for these metal-rich stars,  as expected and known from past studies \citep[e.g.][]{Kordopatis15b, Kordopatis20, GaiaRecio-Blanco23}.

\subsection{Age determination}
\label{sec:ages}
The ages ($\tau$) have been computed using the \citet{Kordopatis23a} pipeline, projecting on PARSEC isochrones \citep{Bressan12}, the XGBoost \teff, \logg, \meta, as well as the extinction-corrected absolute magnitudes in the  $G$, $G_{BP}$, $G_{RP}$, $W_1$, $W_2$ bands. The de-reddening was done 
using the python code {\tt dustmaps} \citep{Green18}, querying  
the \citet{Green19} 3-dimentional map where available, or the \citet{Schlegel98} map, corrected for the stellar distance as in \citet{Kordopatis15b}.  We note that the maximum metallicity reached by the adopted PARSEC isochrones is $+1.0\dex$ and maximum age is $\tau=19.95\Gyr$. 
 From the resulting values, we filter-out the estimations when the age uncertainties are smaller than 0.1$\Gyr$ (i.e. unrealistic uncertainties) or greater than 50 per cent \citep[potentially biased ages, see][]{Kordopatis23a}. 
The  relative age uncertainty  distribution, for the filtered sample, peaks at 28 per cent. 

\begin{figure*}[tbp!]
\begin{center}
\includegraphics[width=0.3\linewidth]{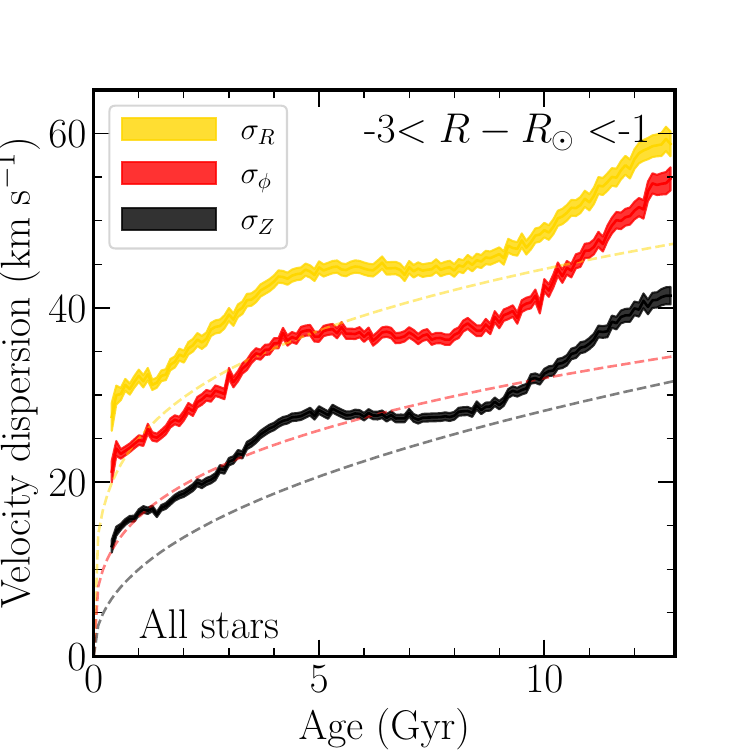}
\includegraphics[width=0.3\linewidth]{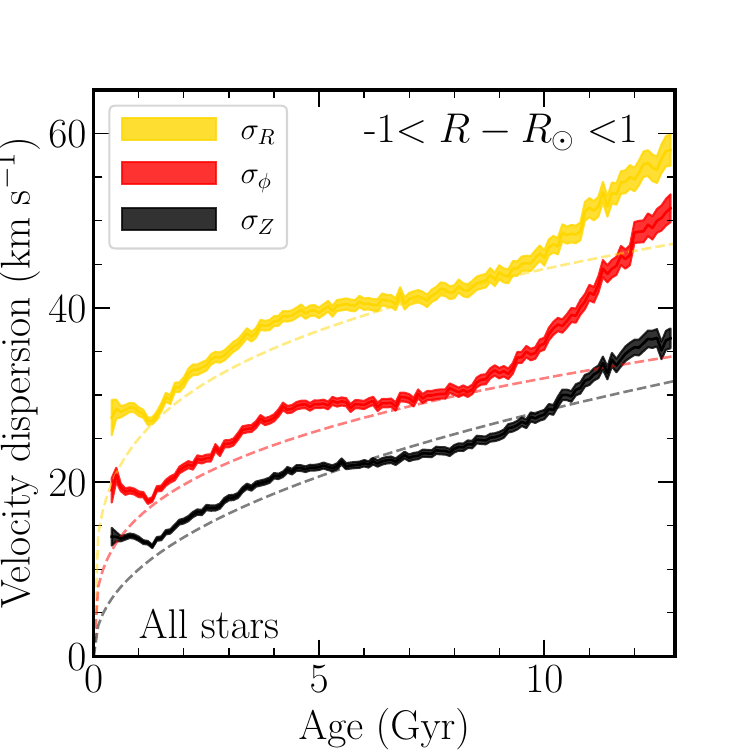}
\includegraphics[width=0.3\linewidth]{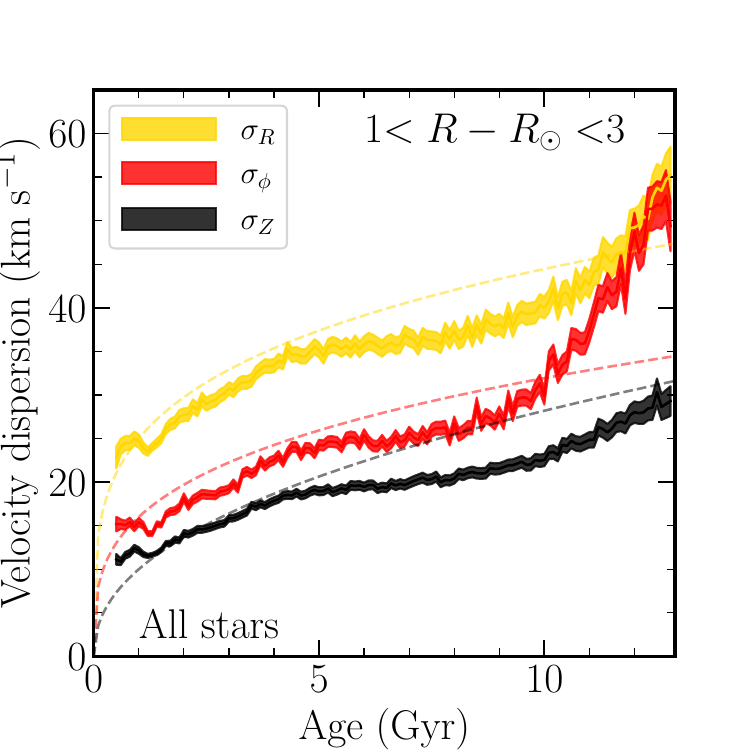}
\caption{Derived age-velocity dispersions for three different 2-kpc wide annuli (Sun being located at $R_{\rm \odot}=8.249\kpc$).  The width of the lines corresponds to the statistical uncertainty on the dispersion ($\sigma/\sqrt{2(n-1)}$, with $n$ the number of stars in a considered bin). Dashed lines are functions in the form of $\sigma_V=\beta \cdot \tau^k$, with adopted values for $\sigma_R, \sigma_\phi, \sigma_Z$ of $\beta=$ [27,18,10] and for $k=$ [0.31,0.34,0.47]. Only stars up to 1\kpc~ from the plane are considered. }
\label{fig:sigma_age_trends}
\end{center}
\end{figure*}

As a validation of our ages, we investigate the derived age-velocity dispersions (in the radial, azimuthal and vertical  directions, noted $\sigma_R, \sigma_\phi$ and $\sigma_z$, respectively) for three different 2-kpc wide annuli, shown in Fig.~\ref{fig:sigma_age_trends}. 
On top of the found trends, we plot the same power-law of the form $\sigma_V=\beta \cdot \tau^k$ as often done in the literature and defined in the Solar neighbourhood \citep[e.g.][see however, \citealt{Seabroke07, Martig14b} for discussions of this not being a sensible choice]{Wielen77, Casagrande11}. We find that our trends can be split in three regimes, a rapid increase (up to 5\Gyr), a rather flat regime (between 5 and 9-10\Gyr) and a jump in the velocity dispersion for the older ages, where the thick disc and the halo start dominating. These are the typical trends expected in our Galaxy, as also found in other studies \citep[see the review of][]{Bland-Hawthorn16}. One can also notice that the $\sigma_{R, \phi, z}$ trends decrease when moving towards larger radii. This will be further discussed in Sect.~\ref{sec:disp_Rg}.

For the stars located at the Solar annulus (middle plot of Fig.~\ref{fig:sigma_age_trends}) we find our derived velocity dispersions for the youngest ages ($\lesssim1\Gyr$)  to be  larger than the exponential law and with a decreasing trend.\footnote{The decrease of the velocity dispersions for the first Gyr in the Solar annulus (middle panel of Fig.~\ref{fig:sigma_age_trends}) is not found when we limit the sample to explicitly the extended Solar neighbourhood, i.e. a bubble with a radius $<1\kpc$ centred on the Sun. } This  maybe due to the accumulation of older stars with underestimated ages. This will be further discussed in the following sections. 
We note, nevertheless, that for $\tau=1.5\Gyr$, we find $(\sigma_R, \sigma_\phi, \sigma_z) \sim (25, 17, 12)\kms$, i.e. in agreement with the  values found in the literature, ranging between 20 and 30$\kms$~for $\sigma_R$, from 12 to 20$\kms$ for $\sigma_\phi$, and from 8 to 15$\kms$ for $\sigma_z$,  see for example \citet{Nordstrom04, Soubiran08, Casagrande11, Mackereth19, McCluskey25}.

\section{Velocity dispersions and age trends across the Galaxy} 
\label{sec:results}

\begin{figure*}[tbp]
\begin{center}
\includegraphics[width=\linewidth]{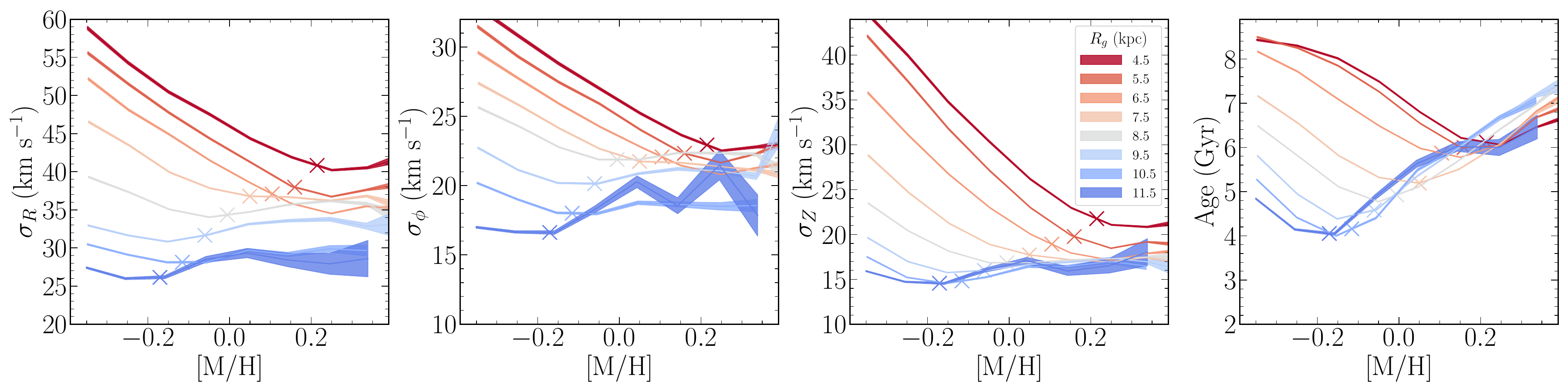}
\caption{Velocity dispersions (radial, azimuthal and vertical in the first, second and third plots, respectively) and average age (fourth plot) as a function of metallicity, for different galactocentric guiding radii $R_g$, as annotated by the legend. The `X' symbols are located at the ISM's metallicity at the considered position, assuming the metallicity gradient derived from classical Cepheids and open clusters from \citet{daSilva23}. The thickness of the line corresponds to the uncertainty on the  measurement. The step in metallicity to compute the trends is 0.1\dex.}
\label{fig:velocity_dispersions}
\end{center}
\end{figure*}

We now investigate the change in the mean age and  velocity dispersions as a function of  metallicity, for different 1\,kpc-wide annuli in the Galaxy. Results are shown in Fig.~\ref{fig:velocity_dispersions}, where the position of the stars is based on their guiding radius. The latter, minimises the effect of blurring by removing the mixing caused by stars on eccentric orbits and hence 
reflect more accurately the mean locus of the stars in the Galaxy.  Figure~\ref{fig:velocity_dispersions_R}, in the appendix, shows the same plots as in Fig.~\ref{fig:velocity_dispersions}, for a selection based on measured galactocentric radius, i.e. $R$. 
On top of each curve, we  plot with an `$\times$' symbol, today's ISM metallicity at the considered location, adopting the \citealt{daSilva23} metallicity $\log$-gradient (see Sect.~\ref{sec:Introduction}).  The $x$-position of the `$\times$'s is therefore completely independent of any measurement done in this study (while the y-position is always our measurement).  Stars with metallicities rightward of this `$\times$' are therefore LMR.

\subsection{Trends as a function of  distance from the Galactic centre}
\label{sec:disp_Rg}
We first focus on the velocity dispersion trends as a function of $R_g$, i.e. at a fixed metallicity. 
Figure~\ref{fig:velocity_dispersions} shows, on average, that the velocity dispersion for all three components drops,  as one moves towards larger Galactic radii (different curves, from red to blue). 

The separation between the velocity dispersion curves gets smaller with increasing $R_g$. This is particularly true  for  $\sigma_z$, for which the trend is  in agreement with \citet{GaiaKatz18, Sanders18} or \citet{Khoperskov24}, and is an additional validation of the adopted  metallicities, positions and kinematics. Indeed  the drop in the velocity dispersion as a function of $R_g$  is  expected for a disc in equilibrium, whose surface density, $\Sigma(R)$, varies as $\sigma_z^2/h_z$, with $h_z$ its scale-height \citep[][]{vanderKruit88}. 
In other words, assuming that $\Sigma(R)$ drops with $R$ \citep[e.g.][]{McMillan17} and that $h_z$ is either constant or increases with $R$ due to flaring \citep[e.g.][]{vanderKruit11, Minchev15}, then $\sigma_Z$ has to decrease with $R$, as we see from our data too (see also Fig.~\ref{fig:sigma_age_trends}).

Regarding the ages,  we find that the stars in the inner Galaxy are older than the stars in the outer Galaxy as long as we consider metallicities below the ones of the ISM. Again, this is the expected trend  for a disc formed inside-out, as we believe it is the case for the Milky Way \citep[e.g.][]{Fall80, Mo98, Frankel19, Minchev19}.
Above the ISM's metallicity, the separation between the age-curves of different $R_g$ is more complex: not only do we find that the separation is tighter, but  it also appears that the age gradient is reversed, i.e. at a fixed metallicity the outer radii have larger ages, on average, than the annuli of smaller $R_g$.

\subsection{Trends as a function of metallicity}
\label{subsection:sigma_vz_metallicity}
\subsubsection{Below the ISM's metallicity}
We now investigate the velocity dispersion and mean age trends as a function of metallicity, i.e. at a fixed $R_g$. 
As mentioned in the previous section, we find, for all velocity components, a drop in $\sigma$ as a function of \meta, up to  \metaism. 
Assuming that, to the $0^{th}$ order, there is an age-metallicity relation (as  confirmed by the fourth plot of Fig.~\ref{fig:velocity_dispersions}, showing an age decrease up to \metaism),
 the drop in $\sigma$ as a function of \meta~ can therefore be associated to the well-known age-velocity dispersion relation in the disc  (see Sect.~\ref{sec:ages}).  Stars with  metallicities lower than the ISM ($\DeltaISM<0$) are either stars born locally that had time to be dynamically heated or stars born at different locations that have been churned and/or blurred, and hence with older ages and hotter kinematics too, compared to locally born stars.

Focusing on $R_g\sim R_\odot$ (grey curve), we find that the velocity dispersion of the stars that have the same metallicity as the  ISM  is  $\sigma_Z\sim16\kms$, i.e. consistent with the literature values of  e.g. \citet{Nordstrom04, Casagrande11, Mackereth19}, for $\tau\sim5\Gyr$ stars, which is also the average age we get from the fourth panel of Fig. ~\ref{fig:velocity_dispersions}.

One could a priori had expected a lower age and a lower velocity dispersion at $R_g= R_\odot$ and \meta=\metaism, since the solar neighbourhood only now reaches solar metallicities for newborn stars, e.g. \citet{Ritchey23}.
 There are at least three factors that can contribute to such high values in our study: 
 
 \emph{ a) A selection bias in our sample against the youngest stars.}
 This point is certainly true. 
 Indeed, no bright massive stars are within the magnitude limits (all stars are fainter than $G\sim9$) and, in addition, the XGBoost catalogue we use is limited to RGB stars. Therefore, young stars are under-represented in the sample, which implies a higher average age in the solar neighbourhood and hence a higher velocity dispersion. This is indeed what we find qualitatively.

\emph{b)  A bias due to the effect of age and metallicity uncertainties.}
 This second point whereas it is true, is not enough to explain the values that we derive. 
 Indeed, Fig.~\ref{fig:sigma_age_trends} indicates that we do not recover properly the expected velocity dispersion for $\tau=0.5\Gyr$ stars ($\sim 14\kms$ instead of $\sim 10\kms$). However, since metallicity uncertainties are expected to be of the order of $\sim0.1\dex$ \citep{Andrae23}, this naturally shifts some metal-poorer stars (and hence older, on average) into higher metallicity bins, inflating that way the velocity dispersion. Appendix~\ref{appendix:contamination}, discusses the effect attributed to a contamination from older and lower-metallicity stars being  part of the high-end tail of the metallicity error distribution, and concludes that if one assumes the reported uncertainties of \citet{Andrae23} as true, then such a contamination cannot explain by itself the large dispersion. 
 
 \emph{c) The presence of  a significant amount of intrinsically old solar metallicity stars in the solar neighbourhood.}
This is also likely true. Several observations of solar-metallicity field stars, with ages determined by different methods corroborate that such stars exist \citep[e.g.][]{Bensby14, Kordopatis15b, Gondoin23, Lehmann24, Nepal24}, whereas the Sun itself is older than $4.5\Gyr$ \citep[e.g.][]{Bahcall95, Bouvier10, Connelly12} and is thought to have migrated by 2-3\kpc~from its birthplace \citep[e.g.][]{Nieva12, Baba23}. 

\subsubsection{Above the ISM's metallicity}
\label{sec:LMR_trends}
 \begin{figure*}[tbp!]
\begin{center}
\includegraphics[width=0.3\linewidth]{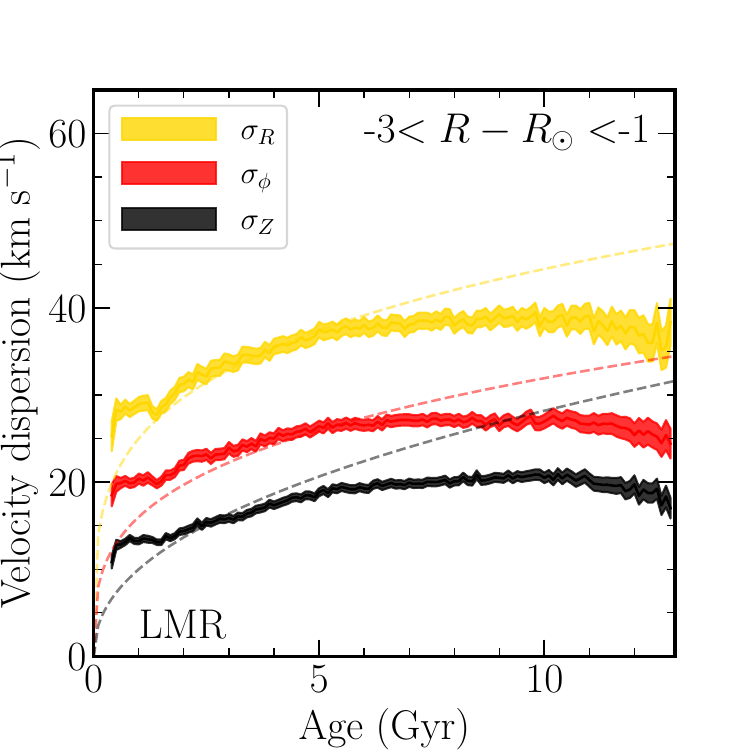}
\includegraphics[width=0.3\linewidth]{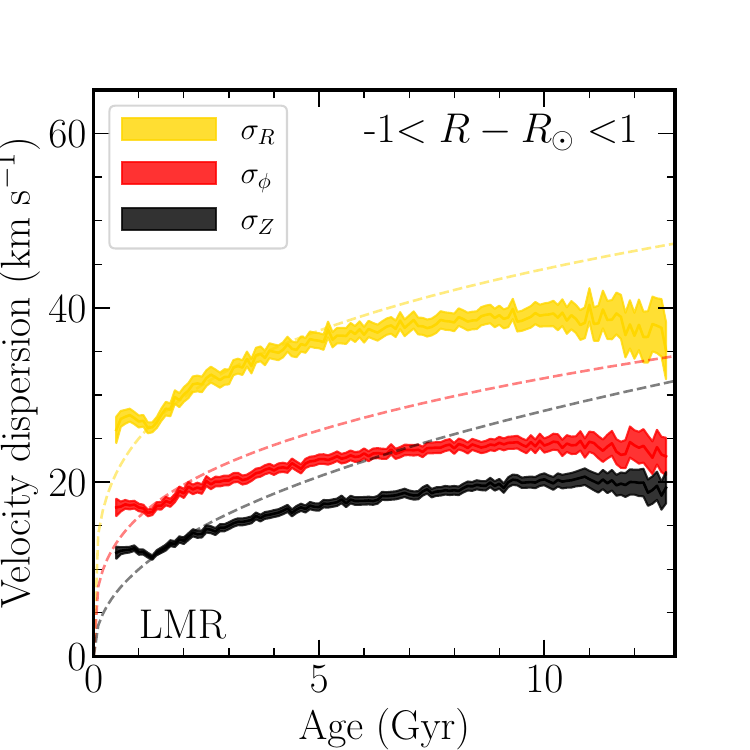}
\includegraphics[width=0.3\linewidth]{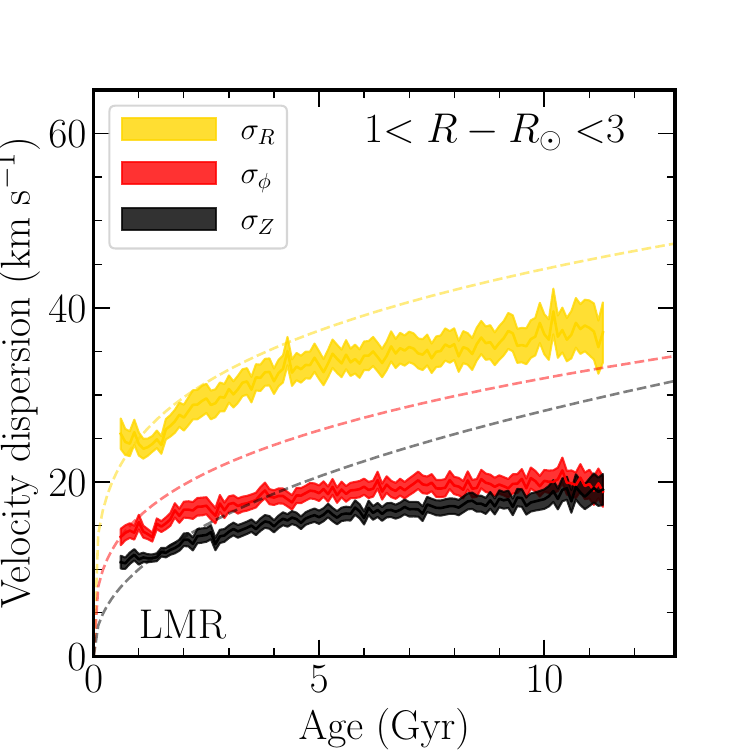}
\caption{Derived age-velocity dispersions for three different 2-kpc wide annuli (Sun being located at $R_{\rm sol}=8.249\kpc$) for local metal-rich stars, i.e. stars that are more metal-rich than the ISM's metallicity at their guiding radius, with $\DeltaISM > 0.1$.  The colours and the dashed lines are same as in Fig. ~\ref{fig:sigma_age_trends}. } 
\label{fig:sigma_age_trends_LMR}
\end{center}
\end{figure*}

For local metal-rich stars, the drop in velocity dispersions as a function of $\meta$, seen for lower metallicities, stops.  That is, the LMR stars have velocity dispersions in all three components  rather flat as a function of metallicity, marginally higher than  the presently locally born stars. This is true for all the considered metallicities above the local ISM,  up to $\DeltaISM\sim+0.5\dex$. 
Indeed, as also shown in Fig.~\ref{fig:sigma_age_trends_LMR}, the velocity dispersions of the LMR stars, as a function of age  is much flatter than the trends found when considering all of the stars (see Fig.~\ref{fig:sigma_age_trends}).

\begin{figure*}[tbp]
\begin{center}
\includegraphics[width=\linewidth]{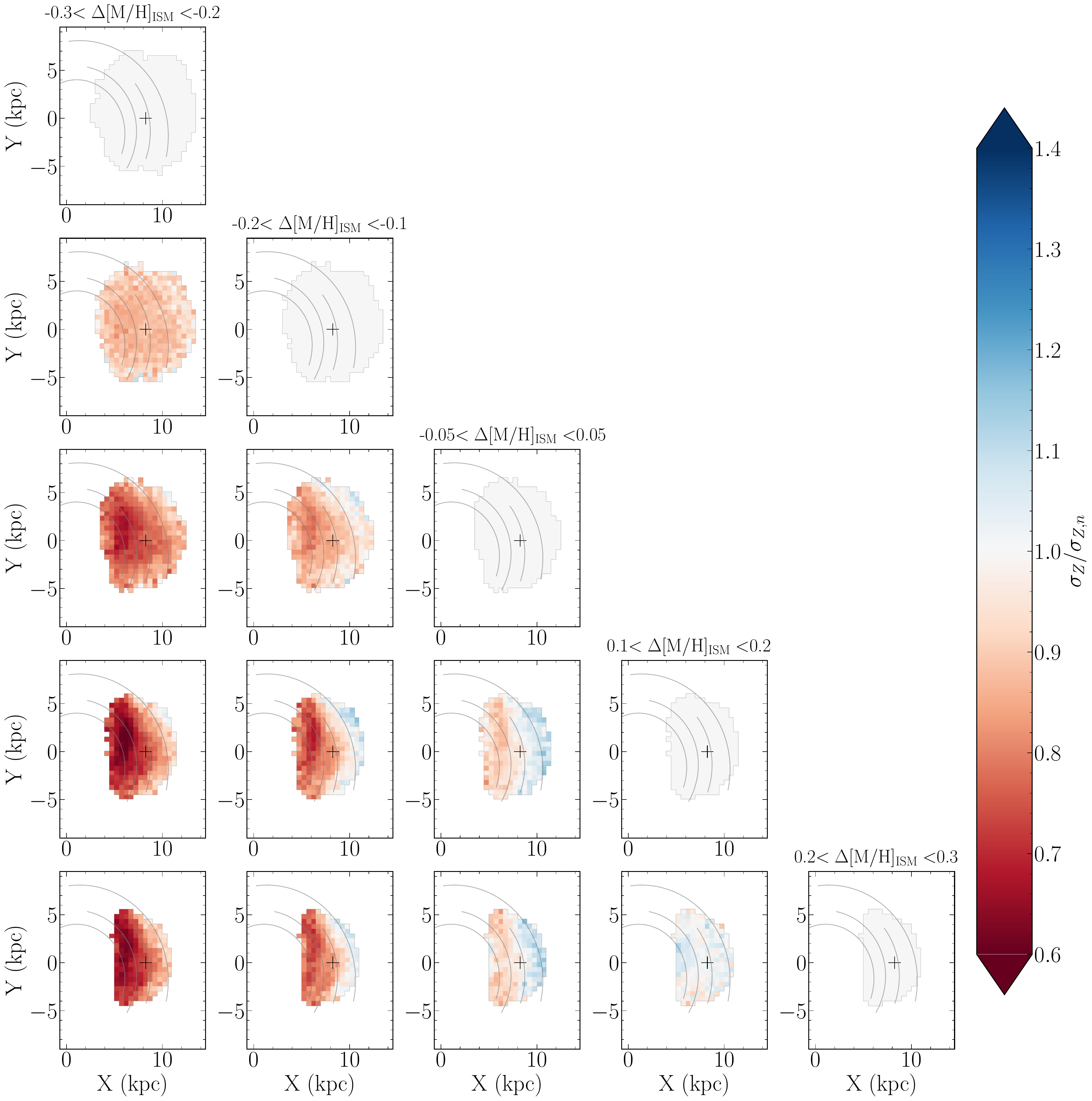} 
\caption{
$X-Y$ maps showing the vertical velocity dispersion ratios between stars at a given metallicity range from the ISM's value (rightmost plots of each row, with the $\sigma_z$ map in grey), and stars of lower metal-content, in bins of 0.1 dex ($\sigma_{Z,n}$, plots leftwards). Over-plotted in grey lines are the spiral arms location, as estimated by \citet{Castro-Ginard21}. Redder (bluer) colours indicate that the lower-metallicity stars have hotter (cooler) kinematics in the considered pixel. 
The assumed Sun's location is indicated by a $`+'$ sign. 
Pixels are 0.5\kpc-wide, and considered only if they include more 200 stars in either selection. Stars with $Z_{\rm max}>2\kpc$ have been discarded.  One can see that at the outermost radii, stars that are considerably more metal-rich than the ISM tend to be slightly kinematically hotter than more metal-poor stars. }
\label{fig:velo_disp_ratio}
\end{center}
\end{figure*}

This trend is global in the disc, and does not depend on the azimuth. This is best shown in the last two rows of Fig.~\ref{fig:velo_disp_ratio}, which illustrates,  in the $X-Y$ plane, the $\sigma_Z$ ratio between stars of different $\DeltaISM$ ranges.  
In this Figure, a clear dichotomy can be noticed for $R\sim8-10\kpc$. Beyond that radius,  LMR stars  of a given $\DeltaISM$ have larger velocity dispersion than stars with metallicities down to $0.2\dex$ below the local ISM (blue colour), whereas this trend reverses at  $R\sim R_\odot$. 
Using the photometric $[\alpha$/M] of \citet{JLi24}, we find no clear evidence that this dichotomy can be due to an `edge' of the high-$\alpha$ disc (which has a larger velocity dispersion), and in particular its absence at large radii \citep[e.g.][]{Bensby11, Kordopatis15a, Hayden15}. 
Relative to the location of the resonances in the disc, this distance, is beyond the bar's co-rotation radius \citep[$\sim6.5-7.5\kpc$, assuming a pattern speed of  $\Omega_b \approx 35-40\kms\kpc^{-1}$, e.g. ][]{Chiba21, Clarke22, GaiaDrimmel23, Zhang24}\footnote{Or even farther, if the pattern speed measured today is the maximum of a fluctuation $\Omega_b$, e.g. \citet{Hilmi20, Li23}. In that case corotation may be even as small as 5\kpc.} and smaller than its outer Lindblad resonance ($R_{\rm OLR} \sim 10.7-12.4$\kpc), but compatible with  the  $m=4$, higher order OLR of $\sim 8.7-10.0\kpc$ \citep{Clarke22}.

 Finally, regarding the age-trends, the right-most panel of Fig.~\ref{fig:velocity_dispersions} shows  a clear uprise of the average age for stars with $\DeltaISM>0$. This rules-out Galactic fountains as the main mechanism forming the LMR, as one may expect either a flatter trend (in case of short radial mixing), or a decrease of the age as a function of metallicity (in the case of large radial mixing, \citealt{Grand19}). This uprise is, in fact,  illustrating that local metal-rich stars need time to migrate and reach their position.
 Furthermore, we find that for all the investigated annuli (except the innermost one) the increase in mean age as a function of metallicity follows a similar trend. This is indicative of the fact that  radial migration is as effective at all the probed radii, hinting towards having the same mechanism causing their churning.

 Another illustration of this can be seen in Fig.~\ref{fig:eccentricity}, which shows the X-Y distribution for stars of different $\DeltaISM$, colour-coded by density, mean eccentricity and mean age, for two different ranges of \zmax.
The last two columns of Fig.~\ref{fig:eccentricity} show clearly, from top to bottom, how we transit from a Galaxy where the inner parts are old and the outer young (first two rows, $|\DeltaISM |<0.1$) to a Galaxy where all of the stars are old ($\DeltaISM >0.1$). Similarly, the larger the  $\DeltaISM$, the older the stars, as they have likely travelled  farther, and needed time to reach their position.

\section{Effect of the spiral arms across the disc on the LMR stars}
\label{sec:effect_of_spirals}
\begin{figure*}[tbp]
\begin{center}
\includegraphics[width=1.\linewidth]{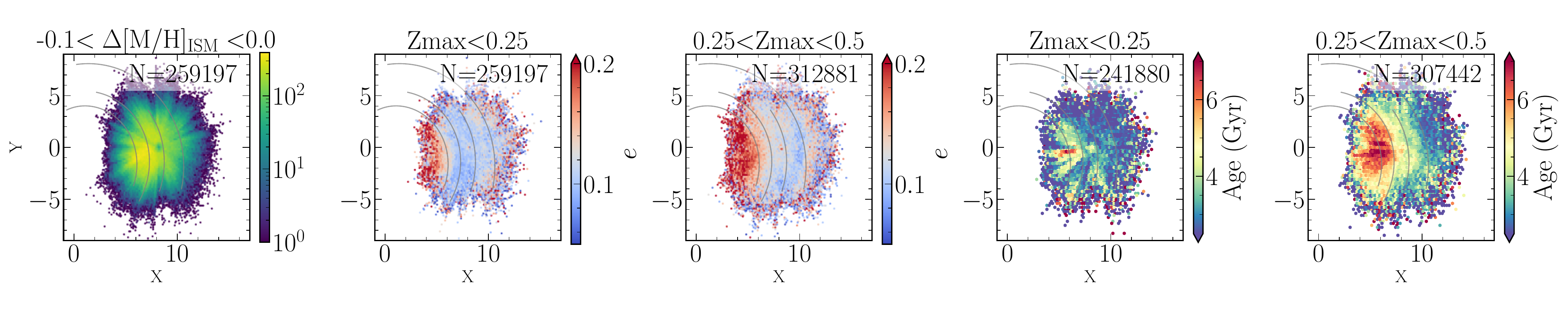}\\ 
\includegraphics[width=1.\linewidth]{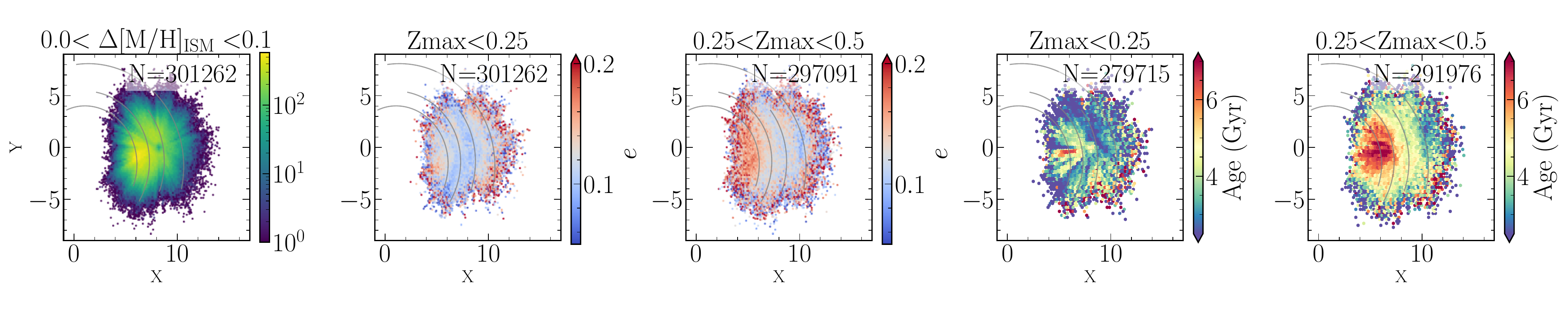}\\ 
\includegraphics[width=1.\linewidth]{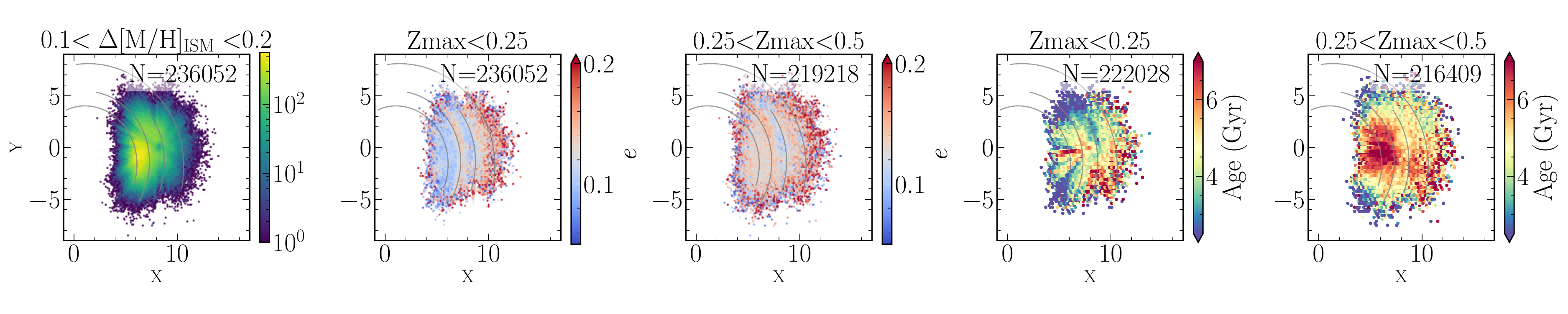}\\
\includegraphics[width=1.\linewidth]{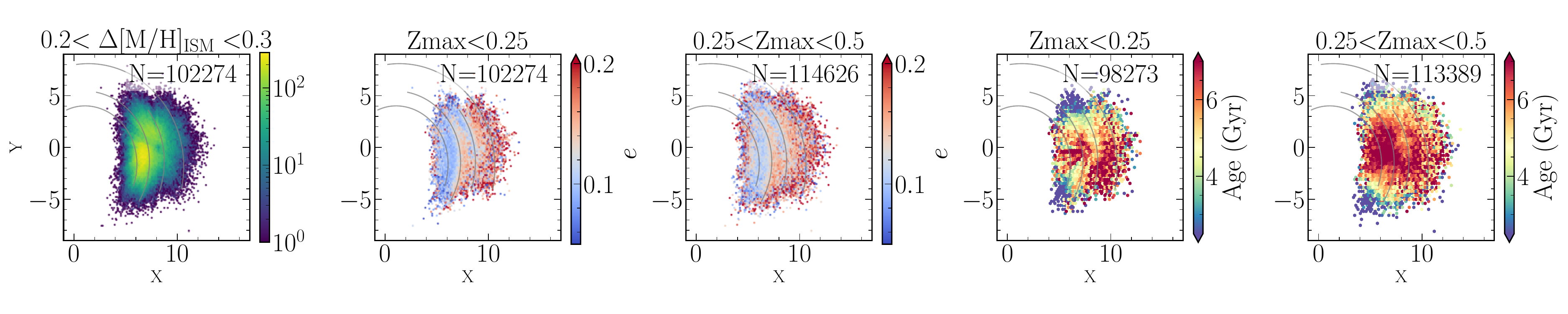} \\
\includegraphics[width=1.\linewidth]{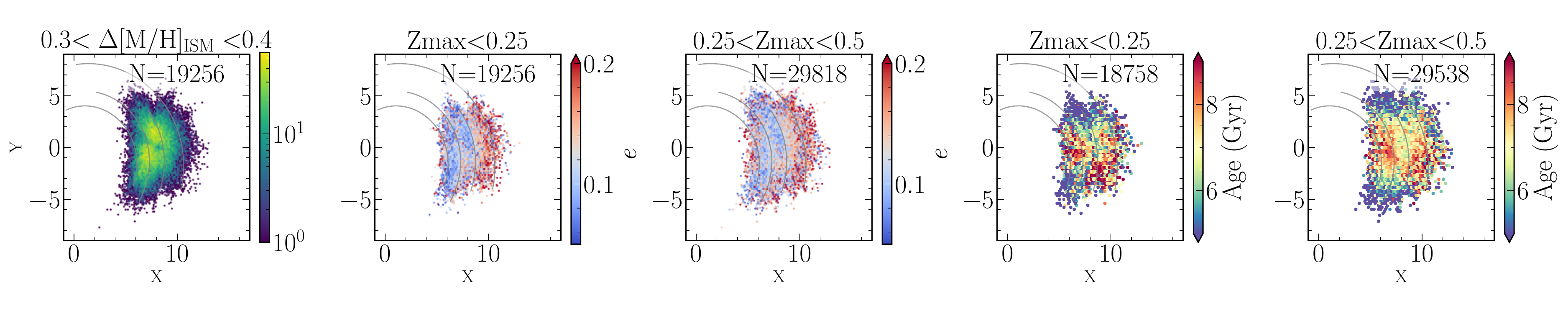} \\
\caption{X-Y maps of stars of different metallicity compared to the ISM's one. Over-plotted in grey lines are the spiral arms location, as estimated by \citet{Castro-Ginard21}. First column: colour-code is density for all the selected stars having \zmax$<0.25\kpc$. Column 2: colour-code is eccentricity, for the stars having a maximum orbital distance from the plane (\zmax) of 0.25\kpc. Column 3: same as column 2, for stars having \zmax~between 0.25 and 0.5\kpc. Columns 4 and 5:  same as columns 2 and 3, colour-coded by age.  We note the age colour-bar for the last row has a different range (from 5 to 9\Gyr) in order to better reflect the older ages, on average, of the extreme LMR stars.
 }
\label{fig:eccentricity}
\end{center}
\end{figure*}

Figures~\ref{fig:velo_disp_ratio} and \ref{fig:eccentricity}  allow us to better comprehend the trends we find with respect to the spiral arms and the central bar. 

We start with the fact that  old LMR stars are found at least up to 11\kpc~from the Galactic centre, i.e. far beyond present day bar's corotation radius (see Sect.~\ref{sec:LMR_trends}). 
Since these stars are mostly on circular orbits ($e\lesssim 0.15$, see columns 2 and 3 of Fig.~\ref{fig:eccentricity}), i.e. they are not visitors on eccentric orbits, we can conclude that the bar alone cannot have shuffled them, and that co-rotation resonances with multiple spiral arms (and potentially overlapping spiral patterns with the bar, see \citealt{Marques25})  need to come into play.

One can also conclude that, at least up to $R\sim$ \Rsun, churned stars do not heat significantly the disc, confirming to some extent the N-body simulation findings of \citet{Minchev12c} or \citet{Halle15}, as well the results obtained with forward-modelling of \citet{Frankel20}. The latter statement, however, needs to be taken with a grain of salt. Indeed, Fig.~\ref{fig:velo_disp_ratio} shows that, farther than \Rsun,   the LMR stars have consistently a velocity dispersion slightly greater ($\sim 20$ per cent) than stars having a metallicity equal to the ISM's. This trend does not seem to be related to the absence of $\alpha$-high stars (i.e. thick disc) at large radii, and cannot be entirely explained by errors in velocity and/or metallicities. This increase, however, may be indicative of the relative quiescent merger history of the Milky Way. 
Indeed, \citet{Minchev12c} have shown that the velocity dispersion decay of migrators ending up at a given radius $R$  can be expressed as $\sigma_{z, \rm mig}(R) \sim \exp{{(-R/4R_d)}}$, which is two times slower than the  one of the non-migrating disc stars  ($\sigma_{z}(R) \sim \exp{{(-R/2R_d)}}$). However, migrating stars usually still end up, with cooler kinematics than local populations either due to a provenance bias \citep[i.e. only dynamically cool stars who spend time in the plane migrate preferentially, see][]{Vera-Ciro14}  and/or  due to the heating of locally born stars by merger interactions \citep{Martig14b}. 
The fact that we find the LMR stars at $R>$\Rsun~ with slightly warmer kinematics seem therefore to indicate that the mergers that have happened during the last 10\Gyr~of the history of the Milky Way have not heated the disc significantly enough, either because they were too few and/or not massive enough \citep[][and references therein]{Deason24}.

 Furthermore, the trend that we measure does not seem to have an azimuthal dependence. This implies that if it is the resonances with the arms that brought the LMR stars where they are, then their non-axisymmetric effects are smoothed out rapidly. 
 
 \bigskip
 
However, we find that the actual presence of the spiral arms does affect, locally, the average eccentricity and age of the LMR stars. Based on Fig.~\ref{fig:eccentricity},  where the spiral arms are plotted in light-grey, as estimated by \citealt{Castro-Ginard21}, one can see that: \\
\smallskip

\emph{a)} the presence of LMR stars does not coincide with over-densities at the location of the spiral arms (column 1),  consistent with spirals being a density wave  \citep{Lin64}, rather than material winding spirals \citep[e.g.][]{Toomre81, Grand12b}. \\
\smallskip

\emph{b)}  LMR stars located at the position of the spiral arms tend to have lower eccentricities (blue stripes in columns 2 and 3 of Fig.~\ref{fig:eccentricity}).The effect of the presence of spirals is seen up to $Z_{\rm max}\sim0.5\kpc$\footnote{Cuts at larger values have been tested and no signatures were found.} and up to metallicities more than 0.4\,dex higher than the ISM.  This is in agreement with \cite{Debattista25, Palicio25} who found that there is a spatial correlation between spiral arms and stellar populations featuring low values of the radial action, and with \citet{Martinez-Medina25} who found the spiral arms extending up to 400\,pc from the plane. \\
\smallskip

\emph{c)} LMR stars located at the position of the spiral arms also show hints of being slightly younger than LMR stars located in-between the arms with the same $\DeltaISM$. This is better visualised in the last two rows, both close and far from the plane, but the uncertainties in the derived ages imply that a thorougher analysis (with, if possible even smaller age-uncertainties) needs to be conducted to confirm this point. \\

\section{Conclusions and perspectives}
\label{sec:conclusions}

In this paper, we have conducted a kinematic and age analysis of the Local Metal-Rich (LMR) stars, i.e. stars whose metallicities exceed that of the interstellar medium (ISM) at their guiding radius. Leveraging Gaia photometric metallicities, we identified and analysed millions of such stars across the Galactic disc.

Our main findings are as follows:
\begin{itemize}
\item
 LMR stars are consistently older than locally born stars with metallicities matching the ISM, yet their velocity dispersions remain only slightly higher, confirming that radial migration does not significantly heat the disc — a key theoretical prediction now supported by observations without modelling.
\item
 At a fixed metallicity excess relative to the ISM, LMR stars located farther out in the disc are systematically older, indicating a time-dependent outward migration consistent with an inside-out growth of the disc and gradual migration over Gyr timescales.

\item
 No correlation is found between spiral arm locations and the density of LMR stars, but the mean stellar eccentricity and age show minima near spiral structures, confirming a kinematic and temporal influence of spiral arms on migrating populations.

\item
The age distribution of LMR stars is inconsistent with a Galactic fountain origin, providing strong evidence in favour of radial migration rather than ejection-reaccretion mechanisms.

\item
 While we do not detect any clear spatial signatures uniquely attributable to the Galactic bar, this does not preclude its influence. As bar-induced migration is expected to be phase-mixed and act over large spatial scales, particularly near resonances like the Outer Lindblad Resonance (OLR), its dynamical effects may not leave obvious azimuthal or spatial imprints in the LMR population. Furthermore, interactions between the bar and spiral arms can generate complex patterns that may be difficult to isolate without detailed dynamical modelling.
\end{itemize}

These results pave the way for more detailed future work. Upcoming large-scale spectroscopic surveys such as 4MOST \citep{deJong19} and WEAVE \citep{Jin24} will deliver more accurate metallicities, stellar ages, and individual elemental abundances. These data will enable refined comparisons with chemical evolution models. Then, using models such as those of \citet{Kordopatis15a} and \citet{Minchev18}, combined with individual elemental abundances tracing different nucleosynthetic channels, it will be possible to place constraints on the birth radii of these stars and to quantify the migration rate as a function of height above the Galactic plane.

 \begin{acknowledgements}
We thank the anonymous referee for their comments and suggestions that helped improving the clarity of the paper. 
GK and VH  gratefully acknowledge support from the French national research agency (ANR) funded project MWDisc (ANR-20-CE31-0004). SF, DF, CL and HE were supported by a project grant from the Knut and Alice Wallenberg Foundation (KAW 2020.0061 Galactic Time Machine, P.I. Feltzing). This project was supported by funds from the Crafoord foundation (reference 20230890). IM acknowledges support by the Deutsche Forschungsgemeinschaft under the grant MI 2009/2-1. DF acknowledges funding from the Swedish Research Council grant 2022-03274. 
Gregor Traven is thanked for the useful plotting tips he gave. Apolline Kordopatis is also thanked for having been a very quiet newborn baby, and Vanessa Gardet for having been an amazing mom during the first weeks of Apolline in this world;   both of them were of an immense help to finalise this paper.  
This work has made use of data from the European Space Agency (ESA) mission
{\it Gaia} (\url{https://www.cosmos.esa.int/gaia}), processed by the {\it Gaia}
Data Processing and Analysis Consortium (DPAC,
\url{https://www.cosmos.esa.int/web/gaia/dpac/consortium}). Funding for the DPAC
has been provided by national institutions, in particular the institutions
participating in the {\it Gaia} Multilateral Agreement.
This research made use of Astropy (\url{http://www.astropy.org}) a community-developed core Python package for Astronomy \citep{astropy:2013, astropy:2018}. 
\end{acknowledgements}

\bibliographystyle{aa}
\def\aj{AJ}\def\apj{ApJ}\def\apjl{ApJL}\def\araa{ARA\&A}\def\apss{Ap\&SS}
\def\mnras{MNRAS}\def\aap{A\&A}\def\nat{Nature}
\def\nar{New Astron. Rev.}

\bibliography{master_bib}

\clearpage

\begin{appendix}
\section{Effect of the different cuts on the metallicity distributions }
\begin{figure}[tbp]
\begin{center}
\includegraphics[width=\linewidth]{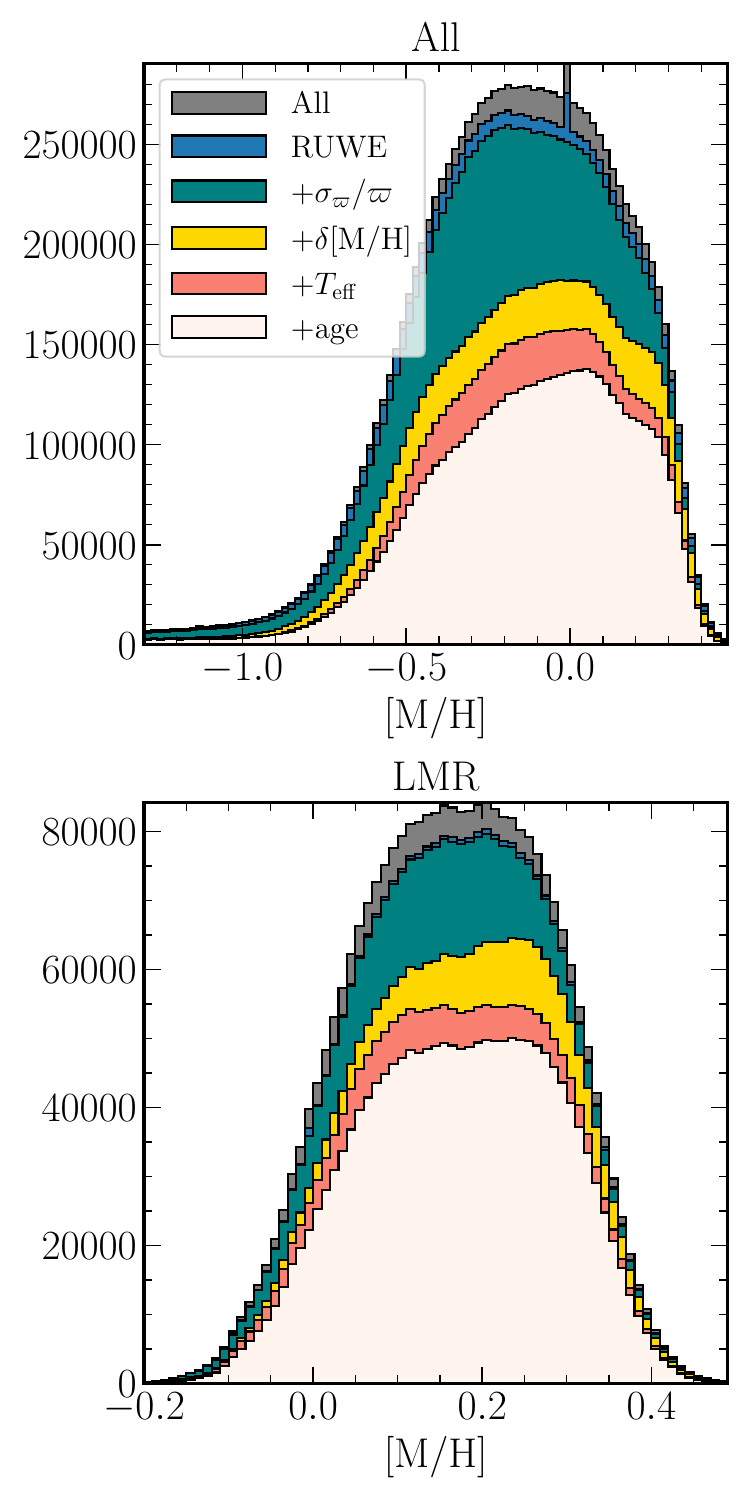}
\caption{Metallicity distribution for the entire RGB sample (top) and the LMR subsample, once the successive quality filters have been applied (see Sect.~\ref{subsec:ruwe_varpi_deltamh_selection}, \ref{sec:teff_filter} and \ref{sec:ages}).}
\label{fig:quality_cuts}
\end{center}
\end{figure}

Figure~\ref{fig:quality_cuts} shows in grey the metallicity distribution between $-1.3$ and $+0.5$ of the XGBoost RGB sample, cross-matched with the AspGap one for all of the stars  ($11\cdot10^6$ stars, top) and the LMR ones ($2.82 \cdot 10^6$, bottom). The effect on the distribution due to the successive filtering on the {\tt RUWE} parameter (in blue), the parallax uncertainty ($\sigma_\varpi/\varpi$, in green), the agreement between the AspGap and XGBoost metallicity ($\DeltaISM$, in yellow) and the \teff~cut (in red)  are also shown (see Sect.~\ref{subsec:ruwe_varpi_deltamh_selection} and \ref{sec:teff_filter}). We note that the parallax uncertainty cut removes most of the spurious XGBoost peak at $\meta=0$. 
Eventually, once all the cuts applied, $\sim 6.36 \cdot 10^6$ stars are left, among which $1.91 \cdot 10^6$ LMR. 
This sample drops to $\sim 6.25 \cdot 10^6$ and $\sim 1.86 \cdot 10^6$ once the filter on the age uncertainty is applied (light pink histogram, see Sect.~\ref{sec:ages}). 

\section{Velocity dispersion at different galactocentric radii}
\begin{figure*}[tbp]
\begin{center}
\includegraphics[width=\linewidth]{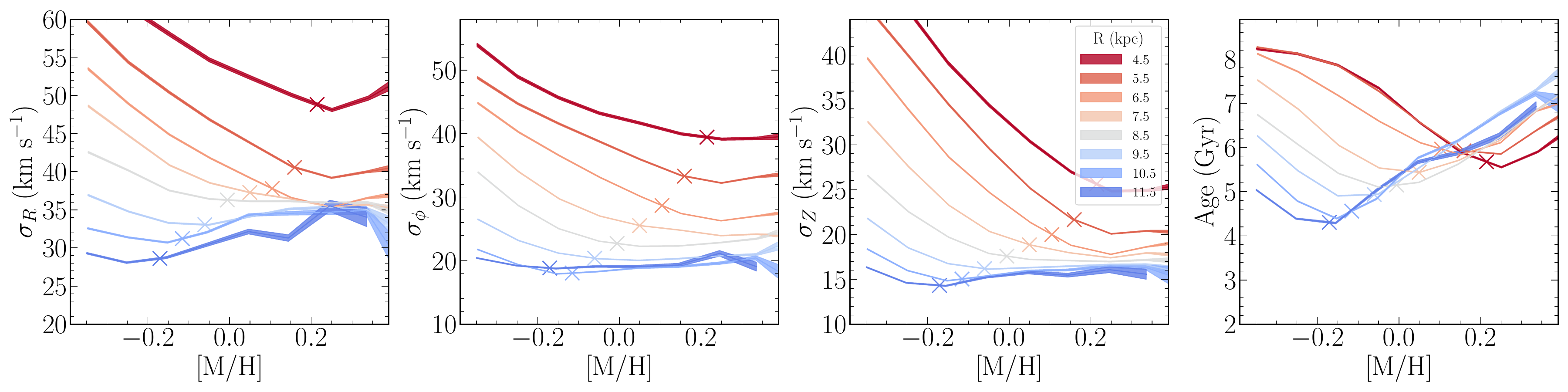}
\caption{Similar to Fig.~\ref{fig:velocity_dispersions}, except that now the position of the stars within an annulus is determined by their actual $R$ position, instead of $R_g$. When plotting as a function of $R$, stars are preferentially seen at their apocenter, therefore on cooler orbits.}
\label{fig:velocity_dispersions_R}
\end{center}
\end{figure*}

Figure~\ref{fig:velocity_dispersions_R} is similar to Fig.~\ref{fig:velocity_dispersions}, where the bins are done in actual Galactic radius, $R$,  rather than guiding radius, $R_g$. Similar velocity dispersion trends as the ones obtained using $R_g$-cuts are found here as well, albeit smoothed out, due to the fact that each $R$-bin can contain stars belonging (due to blurring) to other annuli.

\section{Estimation of the contamination from lower metallicity and older stars}
\label{appendix:contamination}
In Sect.~\ref{subsection:sigma_vz_metallicity} and Fig.~\ref{fig:velocity_dispersions}, we find that $\meta\sim0$ stars at $R_g\sim$\Rsun~ have a vertical velocity dispersion $\sim5\kms$ larger than what found from previous spectroscopic studies. Since kinematic measurements are expected to be very precise because they are based on Gaia data of relatively nearby stars, one can use this value to estimate qualitatively the amount of contamination from older and metal-poorer stars to the considered sample, and see if this contamination is greater than the one expected from  the metallicity uncertainties alone.

Let us consider two populations $A$ and $B$. Population $A$ has a solar metallicity and an age $\tau=0\Gyr$, and population $B$ has $\meta=-0.1$ (i.e. corresponding to the metallicity uncertainty estimated by \citealt{Andrae23}).
Following the age-metallicity relation of e.g.  \citet{Spitoni19}, based on the asteroseimic ages of \citet{Silva-Aguirre18}, stars with $\meta=-0.1$\dex~ can be as old as $5\Gyr$. Adopting an age-velocity dispersion relation in the form $\sigma_z=a \cdot \tau^{k}$ with $a=10\kms$  and $k=0.47$ \citep[e.g.][]{Kordopatis15a}, this results to  $\sigma_{z,B}=21\kms$. 

The combined velocity dispersion, $\sigma_{\rm combined}$, of the two populations, is then defined as  $\sigma_{\rm combined}^2=x\cdot \sigma_A^2 + (1-x)\cdot \sigma_B^2$  with $x$ the proportion of population $A$, and  $\sigma_A$ and $\sigma_B$ are the respective velocity dispersions of each population. 
Since the measured vertical velocity dispersion at $\tau~\sim0.5\Gyr$ in Fig.~\ref{fig:sigma_age_trends} is  $14\kms$, one can conclude that $x=0.75$ and that therefore the contamination of metal-poorer stars is of the order of 25 per cent. 
This value is therefore above the $\sim16$ per cent contamination expected from a unit gaussian centred at $\meta=-0.1$ with a standard deviation of 0.1.

We conclude that, in order to explain the large velocity dispersion we obtain,  one has to assume either that the reported XGBoost uncertainties are underestimated, or that the sample is biased against young stars and/or  contains intrinsically older stars (see discussion in Sect~\ref{subsection:sigma_vz_metallicity}). 

\end{appendix}

\end{document}